\documentclass[10pt]{iopart}
\usepackage{iopams}
\usepackage{epsfig}

\begin{document}
\title{A note on entropy estimation}
\author{Thomas Sch\"urmann}
\address{J\"ulich Supercomputing Centre, J\"ulich Research Centre, 52425 J\"ulich, Germany}

%\ead{thomas.schuermann@live.de}
\vspace{1pc}
\begin{indented}
\item[]\today
\end{indented}

\begin{abstract}
We compare an entropy estimator $\hat{H}_z$ recently discussed in \cite{Zhang} with two estimators $\hat{H}_1$ and $\hat{H}_2$ introduced in \cite{grass03}\cite{schuer}. We prove the identity $\hat{H}_z\equiv \hat{H}_1$, which has not been taken into account in \cite{Zhang}. Then, we prove that the statistical bias of $H_1$ is less than the bias of the ordinary likelihood estimator of entropy. Finally, by numerical simulation we verify that for the most interesting regime of small sample estimation and large event spaces, the estimator $\hat{H}_2$ has a significant smaller statistical error than $H_z$.
\end{abstract}

%\pacs{89.70.+c, 02.50.Fz, 05.45.Tp}
\vspace{1pc}

%\maketitle

\section{Introduction}
Symbolic sequences are typically characterized by an alphabet $A$ of $d$ different letters. We assume statistical stationarity, i.e. any letter-block (word or $n$-gram of constant length) $w_i$, $i=1,...,M$, can be expected at any chosen site to occur with a known probability $p_i=\;$prob$(w_i)$ and $\sum_{i=1}^M\,p_i =1$.

In a classic paper published in 1951, Shannon considered the problem of estimating the entropy
\begin{equation}\label{shan}
H=-\sum\limits_{i=1}^M\,p_i\log p_i,
\end{equation}
of ordinary English \cite{Shannon}. In principle, this might be done by dealing with longer and longer contexts until dependencies at the word level, phrase level, sentence level, paragraph level, chapter level, and so on, have all been taken into account in the statistical analysis. In practice, however, this is quite impractical, for as the context grows, the number $M$ of possible words explodes exponentially with $n$.

In the numerical estimation of the Shannon entropy one can do frequency counting, hence in the limit of large data sets $N$, the relative frequency distribution yields an estimate of the underlying probability distribution. We consider samples of $N$ independent observations, and let $k_i$, $i=1,...,M$, be the frequency of realization $w_i$ in the ensemble. However, with the choice $\hat{p}_i=\frac{\,k_i}{N}$, the naive (or likelihood) estimate
\begin{equation}\label{naiv}
\hat{H}_0=-\sum\limits_{i=1}^M\,\hat{p}_i\log \hat{p}_i
\end{equation}
leads to a systematic underestimation of the Shannon entropy \cite{miller}\cite{harris}\cite{herzel}\cite{gs96}\cite{grass03}\cite{schuer}. In particular, if $M$ is in the order of the number of data points $N$, then fluctuations increase and estimates usually become significantly biased. By bias we denote the deviation of the expectation value of an estimator from the true value. In general, the problem in estimating functions of probability distributions is to construct an estimator whose estimates both fluctuate with the smallest possible variance and are least biased.

On the other hand, there is the Bayesian approach to entropy estimation, building upon
an approach introduced in \cite{Ne02}, or a generalization recently proposed in \cite{Ar14}. There, the basic strategy is to place a prior over the space of probability distributions and then perform inference using the induced posterior distribution over entropy. Actually, a partial numerical comparison of the popular Bayesian entropy estimates and those discussed hereinafter can be found in \cite{Ar14}. Unfortunately, these simulations only consider the bias of the entropy estimates but not their mean square error, which takes into account the important trade-off between bias and variance. However, in the considerations to be discussed below, for what we intend to demonstrate, no explicit prior information on distributions is assumed and we will focus ourself on Non-Bayes entropy estimates only.

To start with, let us consider an estimator of the Shannon entropy which has recently been proposed and analyzed against the likelihood estimator \cite{Zhang}. The development of this interesting estimator starts with a generalization of the diversity index proposed by Simson in 1949 \cite{Simson} and refers to the following representation of the Shannon entropy\footnote{For another interpretation of this representation see \cite{mss}.}
\begin{eqnarray}\label{zz}
H= \sum_{\nu=1}^\infty\frac{1}{\nu}\sum\limits_{i=1}^M p_i \,(1-p_i)^\nu.
\end{eqnarray}
In \cite{Zhang}, it has been mentioned that there exists an interesting estimator of each term in (\ref{zz}), which is unbiased up to the order $\nu=N-1$, namely $Z_\nu/\nu$, where $Z_\nu$ is explicitly given by the expression
\begin{eqnarray}\label{z}
Z_\nu= \frac{N^{1+\nu}(N-\nu-1)!}{N!}\,\sum\limits_{i=1}^M\frac{k_i}{N}
\prod\limits_{j=0}^{\nu-1}\Big(1-\frac{k_i}{N}-\frac{j}{N}\Big)\,,
\end{eqnarray}
such that
\begin{eqnarray}\label{Hz}
\hat{H}_z=\sum_{\nu=1}^{N-1} \frac{1}{\nu}Z_\nu
\end{eqnarray}
is a statistical consistent entropy estimator of $H$ with (negative) bias
\begin{eqnarray}\label{bz}
B_N = -\sum_{\nu=N}^\infty\frac{1}{\nu}\sum\limits_{i=1}^M \, p_i \,(1-p_i)^\nu.
\end{eqnarray}
Indeed, the estimator is notable because a uniform variance upper bound has been proven in \cite{Zhang} that decays at a rate of ${\cal O}(\log(N)/N)$ for all distributions with finite entropy, compared to ${\cal O}((\log(N))^2/N)$ of the ordinary likelihood estimator established in \cite{Antos}. It should be mentioned here that the latter decay rate is an implication of the Efron-Stein inequality, whereas the former (faster) decay rate is derived within the completely different approach introduced in \cite{Zhang}. Actually, it seems hard to prove the same decay rate for the likelihood estimator. \\

In the following section, we will show that $\hat{H}_z$ is algebraically equivalent to the estimator \cite{grass03}\cite{schuer}
\begin{eqnarray}\label{H1}
\hat{H}_1=\sum\limits_{i=1}^M \frac{k_i}{N}\,\Big(\psi(N)-\psi(k_i)\Big),
\end{eqnarray}
while the summation is defined for all $k_i>0$ and the digamma function $\psi(k)$ is the logarithmic derivative of the $Gamma$-function \cite{AS}. Actually, the estimator (\ref{H1}) is given for the choice $\xi=1$ in \cite{schuer} (Eq.\,(28) therein). In the asymptotic regime $k_i\gg 1$ this estimator leads to the ordinary Miller correction $\hat{H}_1\sim\hat{H}_0+(M-1)/2N$. This can be seen by using the asymptotic relation $\psi(x)\sim \log(x)-1/2x$.\\
\\
The mathematical expression of the bias of $\hat{H}_1$ has also been derived in \cite{schuer} and is explicitly given by
\begin{eqnarray}\label{bs1}
B^{(1)}_N = -\sum_{i=1}^M p_i\int_0^{1-p_i} \frac{ t^{N-1}}{1-t}\, dt,
\end{eqnarray}
with a uniform upper bound
\begin{eqnarray}\label{bs2}
|B^{(1)}_N| \le \frac{M}{N}.
\end{eqnarray}
The proof of the identity $B_N\equiv B^{(1)}_N$ will be suppressed here because it is sufficient to show the equivalence of the corresponding entropy estimators in the following section.\\
It should be mentioned that the numerical computation time of the estimator $H_1$ is significantly faster than for $H_z$. Actually, this improvement has not been taken into account in reference \cite{Ar14} ({Fig.\,11}), where the authors still used expression (\ref{Hz}) above. \\
\\
In the third section, by numerical computation we compare the mean square error of $\hat{H}_z$ with an entropy estimator corresponding to $\xi=1/2$ in Eq.\,(13) of \cite{schuer} (see also Eq.\,(35) of \cite{grass03}), which is explicitly given by the following representation
\begin{eqnarray}\label{Hg}
\hat{H}_2=\sum\limits_{i=1}^M \frac{k_i}{N}\Big(\psi(N)-\psi(k_i)+\log(2)+\sum\limits_{j=1}^{k_i-1}\frac{(-1)^j}{j}\Big).
\end{eqnarray}
This estimator is an extension of $\hat{H}_1$ by an oscillating term in the bracket on the right-hand side of (\ref{H1}). In both \cite{grass03} and \cite{schuer}, this estimator has not been expressed in terms of a finite sum, but by integral expressions or infinite sum representations instead. However, it can be easily shown that the present form is equivalent to those in \cite{grass03}\cite{schuer}, but the computation is less time-consuming. The bias of the estimator (\ref{Hg}) is \cite{schuer}
\begin{eqnarray}\label{bs3}
B^{(2)}_N = -\sum_{i=1}^M p_i\int_0^{1-2 p_i} \frac{ t^{N-1}}{1-t}\, dt,
\end{eqnarray}
with uniform upper bound
\begin{eqnarray}\label{bs4}
|B^{(2)}_N| \le \frac{M+1}{2 N}.
\end{eqnarray}
Now, when we look at the right-hand side of (\ref{bs2}) and (\ref{bs4}), then we see that they mainly differ by a factor 2 in the denominator. That has the implication that $|B^{(2)}_N|<|B^{(1)}_N|$ for all $N$ and $M\geq 2$. Thus, we can expect a faster convergence of $\hat{H}_2$ for sufficient large $M$ and not very strongly peaked probability distributions. Actually, these are the distributions we are mainly interested in.
The numerical comparison of the mean square error of $\hat{H}_z$ and $\hat{H}_2$ will be evaluated for the uniform probability distribution, the Zipf distribution and for the zero-entropy delta distribution.

\section{Comparison of $\hat{H}_z$ and $\hat{H}_1$}

In this section, we show the identity $\hat{H}_z\equiv\hat{H}_1$. Therefore, let $Z_{i,\nu}$ denote the $i$-$th$ term of (\ref{z}),
\begin{eqnarray}\label{zi1}
Z_{i,\nu}= \frac{N^{1+\nu}(N-\nu-1)!}{N!}\,\frac{k_i}{N}\,\prod\limits_{j=0}^{\nu-1}(1-\frac{k_i}{N}-\frac{j}{N}).
\end{eqnarray}
By extending with $N$ in the product, this expression can be rewritten as
\begin{eqnarray}\label{zi2}
Z_{i,\nu}= \frac{(N-\nu-1)!}{N!}\,k_i\,\prod\limits_{j=0}^{\nu-1}(N-k_i-j).
\end{eqnarray}
Next, the product is reformulated as a quotient of factorials, i.e.
\begin{eqnarray}\label{f}
\prod\limits_{j=0}^{\nu-1}(N-k_i-j) = \frac{(N-k_i)!}{(N-k_i-\nu)!}
\end{eqnarray}
and in terms of binomial coefficients we get
\begin{eqnarray}\label{zi4}
Z_{i,\nu}= \frac{k_i}{N-\nu}\,{N-\nu\choose k_i}\Big/{N\choose k_i}.
\end{eqnarray}
Now, the $i$-$th$ term of the estimator (\ref{Hz}) is obtained by summation over $\nu$, i.e.
\begin{eqnarray}\label{zi5}
\sum_{\nu=1}^{N-1} \frac{1}{\nu}\,Z_{i,\nu}
&=& {N\choose k_i}^{-1}k_i\,\sum_{\nu=1}^{N-1} \frac{1}{\nu(N-\nu)}\,{N-\nu\choose k_i}\nonumber\\
&=& \frac{k_i}{N}\,\Big({\cal H}_{N-1}-{\cal H}_{k_i-1}\Big)
\end{eqnarray}
while ${\cal H}_k=\sum_{n=1}^k 1/n$ is the $k$-$th$ harmonic number \cite{AS}. Applying the identity ${\cal H}_{k-1}=\psi(k)+\gamma$ (with $\gamma=0.5772...$, the Euler{-}Mascheroni constant) and summation for $i=1,2,...,M$, we obtain the estimator (\ref{H1}), which proves the identity $\hat{H}_z\equiv\hat{H}_1$.\\
\\
In addition, we have the following\\
\\
$\mathbf{Proposition.}$ The estimator $\hat{H}_1$ is less biased than the likelihood estimator $\hat{H}_0$. \\
\\
$\mathbf{Proof.}$
Since we know from \cite{schuer} that the bias of $\hat{H}_1$ is negative, it is sufficient to prove that $\psi(N)-\psi(k)> \log\frac{N}{k}$, for $0<k<N$. The following inequalities \cite{AS}
\begin{eqnarray}
\psi(N)&\ge& \log\big(N-\frac{1}{2}\,\big)\\
\psi(k)&\le& \log(k) -\frac{1}{2k}
\end{eqnarray}
can be applied such that we only have to check that
\begin{eqnarray}\label{in}
N>\frac{1/2}{\;1-e^{-\frac{1}{2k}}}.
\end{eqnarray}
Now, for any finite $k>0$, the inequality $1+\frac{1}{2k}<\exp\left(\frac{1}{2k}\right)$ is satisfied. The proof is by Taylor series expansion of the exponential function. From this, by simple algebraic manipulations, it follows that the right-hand side of (\ref{in}) is less than $k+\frac{1}{2}$, for any finite $k>0$. It follows that (\ref{in}) is satisfied for any $k$ with $0<k<N$. This proves that $\hat{H}_1$ is less biased then $\hat{H}_0$.$\hfill\Box$

\section{Numerical comparison of $\hat{H}_z$ and $\hat{H}_2$}

In this section, we will focus on the convergence rates of the root mean square error (RMSE) of $\hat{H}_z$ and $\hat{H}_2$. Here, the RMSE is defined by
\begin{eqnarray}\label{mse}
\textrm{RMSE}=\sqrt{\textrm{E}[(\hat{H}-H)^2]}.
\end{eqnarray}
We choose this error measure because it takes into account the trade-off between bias and variance. Moreover, we want to mention that there is a slightly modified version $\hat{H}^*_z$ of the estimator $\hat{H}_z$,  defined in Eq.\,(12) of \cite{Zhang}. Since the bias $B_N$ of $\hat{H}_z$ is explicitly known, a correction is defined by subtraction of the bias term $B_N$ with $p_i$ replaced by its estimate $\hat{p}_i$. The modified estimator is then given by $\hat{H}^*_z=\hat{H}_z-\hat{B}_N$, while $\hat{B}_N$ is the plug-in estimator of $B_N$. For reasons of simplicity, we deny applying the same procedure of bias correction for the estimator $\hat{H}_2$.\\

\begin{figure}[t]
\begin{center}
\psfig{file=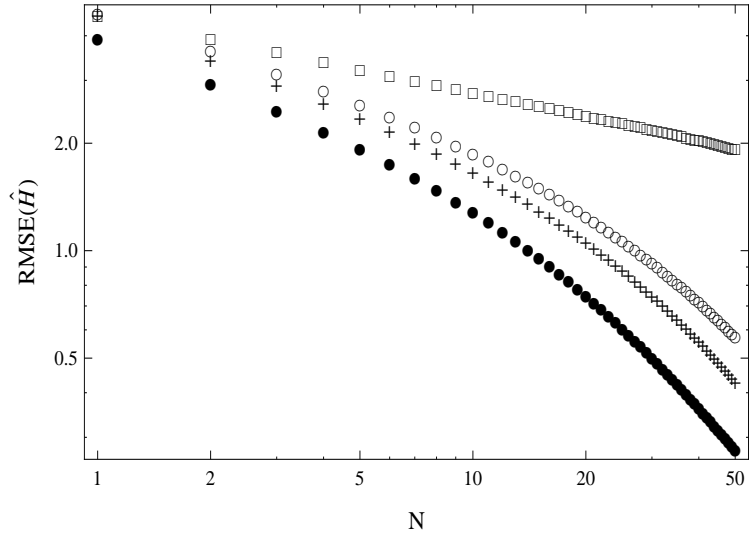,width=10.0cm, height=7.5cm}
\end{center}
\parbox{11.5cm}{
\caption{Statistical error of $\hat{H}_0$ ({\scriptsize $\square$}), $\hat{H}_z$ ($\circ$), $\hat{H}^*_z$ (+) and $\hat{H}_2$ ($\bullet$), for the uniform probability distribution with $M=100$ (see text). The RMSE of $\hat{H}_2$ is significantly smaller then of $\hat{H}_z$ and $\hat{H}^*_z$. The exact value of the entropy is $H=5.3$.\label{fig1} }
}
\end{figure}

\begin{figure}[t]
\begin{center}
\psfig{file=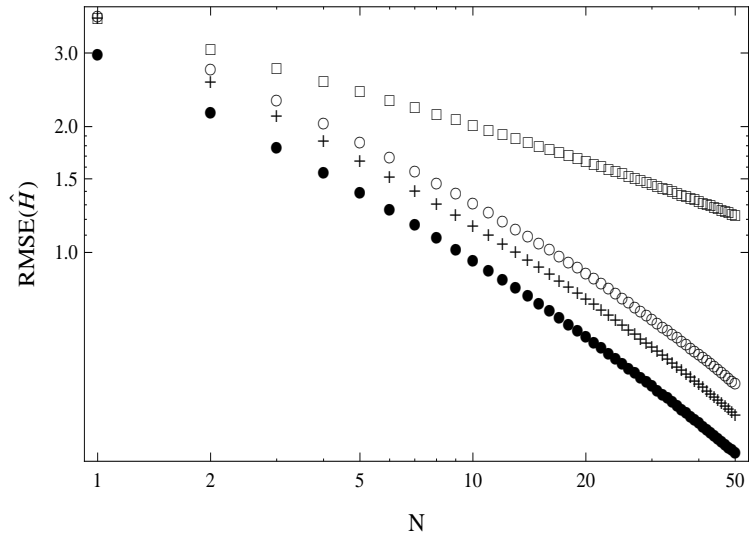,width=10.0cm, height=7.5cm}
\end{center}
\parbox{11.5cm}{
\caption{Same as in Fig.\,\ref{fig1}, but for Zipf's probability distribution (see text). The exact value of the entropy is $H=3.68$.\label{fig2}}
}
\end{figure}

Our first data sample is taken from the uniform probability distribution $p_i=1/M$, for $i=1,2,...,M$. In addition, we consider the (right-tailed) Zipf-distribution with $p_i=c/i$, for $i=1,2,...,M$ and normalization constant $c=1/{\cal H}_M$ (reciprocal of the $M$-$th$ harmonic number). The statistical error for increasing sample size $N$ and given $M$ is shown in Fig.\,\ref{fig1} and Fig.\,\ref{fig2}. As we can see, the RMSE of all estimators is monotonic decreasing in $N$. The convergence of the naive estimator $\hat{H}_0$ is rather slow compared to the other estimators, while the performance of $\hat{H}^*_z$ is slightly better than for $\hat{H}_z$. On the other hand, the statistical error of $\hat{H}_2$ is significantly smaller than the statistical error of $\hat{H}_z$ and $\hat{H}^*_z$ and this behaviour seems to be representative for large $M$.

\begin{figure}[t]
\begin{center}
\psfig{file=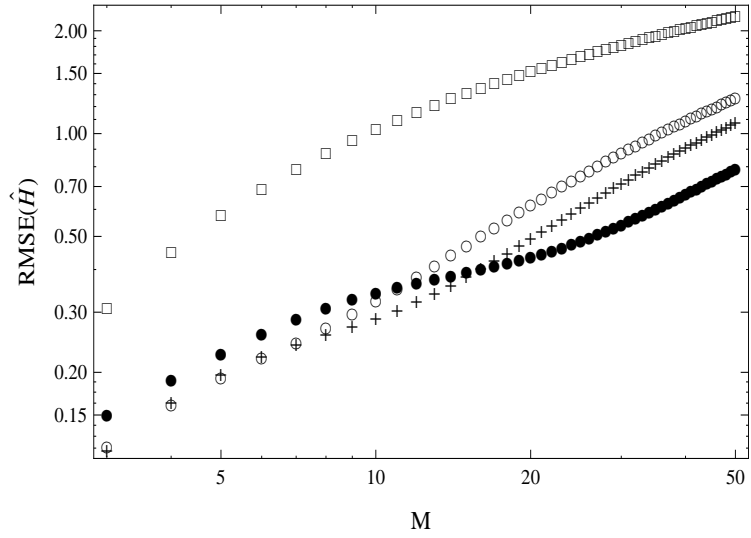,width=10.0cm, height=7.5cm}
\end{center}
\parbox{11.5cm}{
\caption{Statistical error of $\hat{H}_0$ ({\scriptsize $\square$}), $\hat{H}_z$ ($\circ$), $\hat{H}^*_z$ (+) and $\hat{H}_2$ ($\bullet$), for sample size $N=10$ in the instance of the uniform probability distribution. Small sample estimation is expected when $M$ is above the sample size $N$.\label{fig3}}
}
\end{figure}

\begin{figure}[t]
\begin{center}
\psfig{file=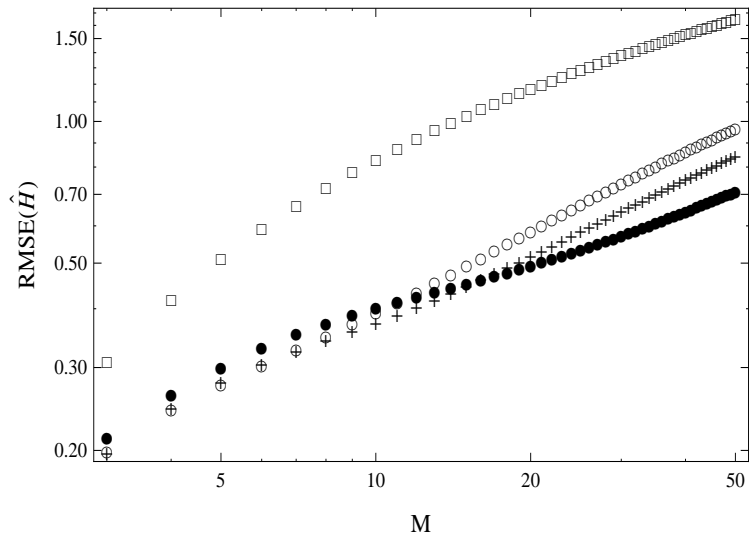,width=10.0cm, height=7.5cm}
\end{center}
\parbox{11.5cm}{
\caption{Same as in Fig.\,\ref{fig3}, but for the Zipf distribution. There is a crossover for $M\approx N$. \label{fig4}}
}
\end{figure}

The statistical error for increasing $M$ and fixed sample size $N$ is shown in Fig.\,\ref{fig3} and Fig.\,\ref{fig4}. For $M\gg N$, the RMSE of $\hat{H}_z$ and $\hat{H}^*_z$ is greater than of $\hat{H}_2$. This phenomenon reflects the fact that the bias reduction becomes more and more relevant for increasing $M$, compared to the contribution of the variance.

As we can see from both examples, the gap between $\hat{H}^*_z$ and $\hat{H}_2$ is slightly smaller for the peaked Zipf distribution compared to the uniform distribution. Thus, we ask for the performance in the extreme case of the delta distribution $p_i=\delta_{i,1}$, which has entropy zero. Indeed, in this special case we have $\hat{H}_0=\hat{H}_1=\hat{H}_z=\hat{H}^*_z= 0$ for any sample size $N$, but $\hat{H}_2=\log(2)+\sum_{j=1}^{N-1}(-1)^j/j\to 0$ for $N\to\infty$. Actually, in this case the statistical error of the latter scales like $\sim 1/2N$ for large $N$.

\section{Summary}

In the present note, we classified the entropy estimator $\hat{H}_z$ of \cite{Zhang} within the family of entropy estimators originally introduced in \cite{schuer}. This reveals an interesting connection between two different approaches to entropy estimation, one coming from the generalization of the diversity index of Simpson and the other one coming from the estimation of $p_i^q$ in the family of Renyi entropies. This connection is explicitly established by the identity $\hat{H}_z\equiv \hat{H}_1$. In addition, we proved that the statistical bias of $\hat{H}_1$ is smaller than the bias of the likelihood estimator $\hat{H}_0$.

Furthermore, by numerical computation for various probability distributions, we found that $\hat{H}_z$ (or the heuristic estimator $\hat{H}^*_z$) can be improved by the estimator $\hat{H}_2$, which is an excellent member of the estimator family in \cite{grass03}\cite{schuer}.

On the other hand, there is a uniform variance upper bound of $\hat{H}_z$ (and therefore of $\hat{H}_1$) that decays at a rate of ${\cal O}(\log(N)/N)$ for all distributions with finite entropy \cite{Zhang}. It would be interesting to know if this variance bound also holds for the estimator $\hat{H}_0$ or $\hat{H}_2$. The answer might be found in a forthcoming publication.\\

%\newpage
{\bf References}\\

\newpage

\end{document}